\documentclass[prb,twocolumn,showpacs,preprintnumbers,amsmath]{revtex4}
\usepackage{graphicx}
\usepackage{dcolumn}

\begin{document}

\title{Pseudogap and Mott Transition Studied by Cellular Dynamical Mean Field Theory}
\author{Y. Z. Zhang and Masatoshi Imada}
\affiliation{Department of Applied Physics, School of Engineering, University of Tokyo, Hongo 7-3-1, Bunkyo-ku, Tokyo 113-8656, Japan}
\date{\today}

\begin{abstract}
We study metal-insulator transitions between Mott insulators and metals. The transition mechanism completely different from the original dynamical mean field theory (DMFT) emerges from a cluster extension of it.  A consistent picture suggests that the quasiparticle weight $Z$ remains nonzero through metals and suddenly jumps to zero at the transition, while the gap opens continuously in the insulators. This is in contrast with the original DMFT, where $Z$ continuously vanishes but the gap opens discontinuously. The present results arising from electron differentiation in momentum space agree with recent puzzling bulk-sensitive experiments on CaVO$_3$ and SrVO$_3$. 
\end{abstract}

\maketitle

Mechanisms and nature of correlation-induced metal-insulator transitions (MIT) are fundamental challenges in condensed matter physics\cite{Imada}. 
Although the metals and insulators far away from the MIT are relatively well understood, electronic states dominated and controlled by the proximity of the MIT are far from complete understanding. We find many challenging phenomena such as the high-$T_c$ superconductivity in this region, which waits for better understanding of underlying proximity of MIT.   

The dynamical mean-field theory (DMFT) offers a clear picture of the MIT by taking the limit of high dimensions~\cite{Metzner}. 
In the DMFT, the MIT is approached from metals by the reduction of the quasiparticle weight $Z$ at the Fermi level and the transition is driven by vanishing $Z$~\cite{Georges}.  At the transition, a nonzero insulating gap is already open in the density of states (DOS) as a result of the vanishing $Z$ of the coherent peak isolated from the well-separated upper and lower Hubbard bands. Early photoemission spectroscopy (PES) and inverse-PES of Ca$_{1-x}$Sr$_x$VO$_3$~\cite{Sekiyama} showed a three peak structure, called a sharp coherent band at $E_F$ and the incoherent ones corresponding to the upper and lower Hubbard bands at a few electron volts above and below $E_F$, which is consistent with the above DMFT scenario. However, recent more bulk-sensitive PES have revealed a qualitatively different and puzzling feature with a new broad peak at around 0.2eV connected to the lower Hubbard band around 1.6eV and a pseudogap formation around the Fermi level~\cite{Eguchi}. This is in marked contrast with the DMFT results.

In this paper, we show that a scenario of the MIT completely different from the original DMFT emerges in two and three dimensional systems by extending the DMFT to allow the momentum dependence of the self-energy. The MIT is now not driven by a continuous reduction of $Z$ to zero, but by the quasiparticle poles moving away from the Fermi surface, leading to a discontinuous jump of $Z$ to zero at the Fermi level. Such a MIT is the consequence of an anisotropic extinction of the Fermi surface with the topological change, namely the electron differentiation in momentum space. We show the results by considering a cluster extension of the DMFT up to four sites for the Hubbard model on the square and cubic lattices. Our results indicate an opening of pseudogap in the DOS already in the metallic side leading to the continuous opening of the true gap in the insulator at a finite energy resolution. This offers a picture consistent with the recent bulk-sensitive PES~\cite{Eguchi}.

In spite of the substantial success, limitations of the original (single-site) DMFT due to the ignorance of the momentum dependence in the self-energy have been repeatedly pointed out~\cite{Maier,Kotliar}. Recently, such as cellular DMFT (CDMFT)~\cite{Kotliar2}, dynamical cluster approximation~\cite{Hettler}, cluster perturbation theory~\cite{Senechal}, self-energy functional approach~\cite{Potthoff} and correlator projection theory~\cite{Onoda} have been developed with some promising results on the two-dimensional Hubbard model by focusing on either doping or temperature effect~\cite{Maier,Senechal2,Senechal3,Kyung,Moukouri,Huscroft,Parcollet}. Among these results, only Kyung, et. al.~\cite{Kyung} first mentioned a four-band structure at half filling as the Mott transition is approached, which is remarkably different from the band structure predicted from the standard DMFT~\cite{Georges}. However, since no further analysis was performed, such as the analysis of the k-dependence of quasiparticle weight, self energy and spectral density, as well as the correlation functions, the mechanism and the criticality of the MIT cannot be derived from their results, while the new PES~\cite{Eguchi} casts doubts on the widely accepted MIT scenario~\cite{Georges} and requires a new description of the MIT. Moreover, it remains unclear if the choice of cluster size and the dimensionality will affect the results. In this paper, we study in detail the two- and three-dimensional Hubbard model with the transfers for the nearest neighbor pairs with the standard notation:
\begin{equation}
H=-t\sum_{\left\langle i,j\right\rangle ,\sigma }(c_{i\sigma }^{\dagger
}c_{j\sigma }+\mathrm{H.c.})+U\sum_in_{i,\uparrow }n_{i,\downarrow }.  \label{Hubbard}
\end{equation}
on square and cubic lattices at half-filling with CDMFT. The CDMFT has aleady passed several tests against the exact results and the density matrix renormalization group method in one dimension\cite{Bolech}, where the CDMFT scheme is expected to be worst. We employ exact diagonalization (ED) as an impurity solver of CDMFT\cite{Caffarel}. The advantage of ED is the access to zero temperature and real frequency, as well as the ability to treat the large interaction regime, which is hardly accessible by the quantum Monte Carlo method.

Throughout this paper, we restrict our studies in the paramagnetic state. The magnetic or charge density wave order are indeed absent in SrVO$_3$ and CaVO$_3$. We take into account the temperature effect as well as the finite energy resolution by introducing a Lorentzian broadening factor. Finally, we should point out that, the CDMFT always underestimate the $k$-dependent behavior of self-energy  owing to the limitation of cluster size. Therefore the possible non-Fermi-liquid behavior at asymptotically low-energy scale is hardly tractable. Nevertheless, even from our small cluster CDMFT study, we find that the scenario of MIT from DMFT  breaks down. 

To implement the CDMFT approach, it is convenient to consider the cluster effective action derived from a cluster-impurity Hamiltonian of the form 
\begin{eqnarray}
&&H_{CI}=\sum_{\left\langle \mu \nu \right\rangle \sigma }t_{\mu \nu }c_{\mu
\sigma }^{\dagger }c_{\nu \sigma }+U\sum_\mu n_{\mu ,\uparrow }n_{\mu
,\downarrow }  \nonumber \\
&&+\sum_{m\sigma }\varepsilon _{m\sigma }a_{m\sigma }^{\dagger }a_{m\sigma
}+\sum_{m\mu \sigma }V_{m\mu \sigma }^h(a_{m\sigma }^{\dagger }c_{\mu \sigma
}+\mathrm{H.c.}),  \label{Impurity}
\end{eqnarray}
where the indices $\mu ,\nu =1,...,N_c$ label sites within the cluster, while $m=1,...,N_b$ represent the bath degrees of freedom, and $c_{\mu\sigma }$ ($a_{m\sigma }$) annihilates an electron on the cluster (bath), respectively. $t_{\mu \nu }$ is the hopping matrix within the cluster, $\varepsilon _{m\sigma }$ is the bath energy and $V_{m\mu \sigma }^h$ is the bath-cluster hybridization matrix. We begin with an initial guess for the bath parameters $\varepsilon _{m\sigma }$ and $V_{m\mu \sigma }^h$ determining an initial cluster Weiss field $G_{0c}^\sigma \left( i\omega \right) $. With this guess the cluster Green's function (GF) $G_c^\sigma \left( i\omega \right) $ is calculated by solving the cluster-impurity Hamiltonian (\ref{Impurity}). Applying the Dyson equation, $\Sigma _c^\sigma =\left( G_{0c}^\sigma \right) ^{-1}-\left(G_c^\sigma \right) ^{-1}$, we obtain the cluster self-energy. To close the self-consistency loop, we obtain a new $G_{0c}^{\sigma \prime }\left( i\omega\right) $ from 
\begin{eqnarray}
&(&G_{0c}^{\sigma \prime }\left( i\omega _n\right) )^{-1}-\Sigma _c^\sigma
\left( i\omega _n\right)  \label{selfconsistent} \\
&=&\left( \frac{N_c}{\left( 2\pi \right) ^D}\int \frac{d\widetilde{k}}{%
i\omega _n+\mu -t\left( \widetilde{k}\right) -\Sigma _c^\sigma \left(
i\omega _n\right) }\right) ^{-1}.  \nonumber
\end{eqnarray}
Here $t\left( \widetilde{k}\right) $ is the hopping matrix for the superlattice with the wavevector $\widetilde{k}$ because of the inter-cluster hopping and $D$=2 for square lattice, 3 cubic lattice. The bath parameters for the next iteration is determined by minimizing the following distance function 
\begin{equation}
d=\sum_{\omega _n,\mu ,\nu }\left| \left[ \left( G_{0c}^{\sigma \prime
}\left( i\omega _n\right) \right) ^{-1}-\left( G_{0c}^{\sigma ,N_b}\left(
i\omega _n\right) \right) ^{-1}\right] _{\mu \nu }\right| ^2.
\label{distant}
\end{equation}

\begin{figure}[tb]
\begin{center}
\includegraphics[width=8.0cm]{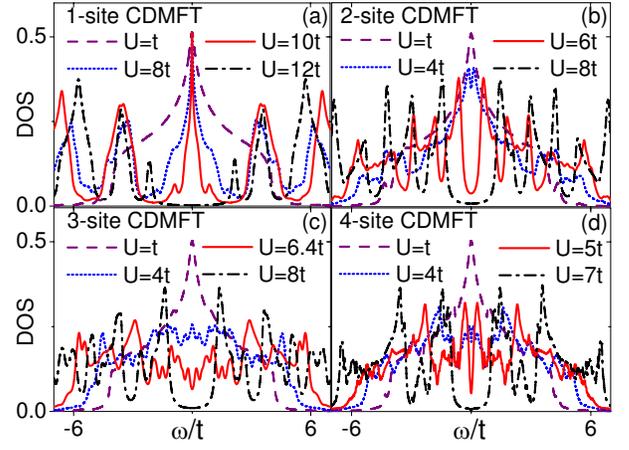}
\end{center}
\caption{(Color online) Density of state of Hubbard model on half-filled square lattice at four different values of $U$, obtained from CDMFT with cluster size up to 4. (a) is the 1-site CDMFT which reproduces the results of the single-site DMFT. The central coherent peaks are narrowed but persist until the phase transition. (b),(c),(d) are 2-, 3-, and 4-site CDMFT, respectively. In all the three cases, the central peak is suppressed while a twin-peak structure appears around Fermi level. We take the broadening factor 0.1$t$, viewed as a temperature effect or a finite energy resolution in experiments, a pseudogap is observed prior to the Mott transition.}
\label{Fig.1}
\end{figure}

In the present study, we choose the size of cluster from 1 up to 4. For 1-site CDMFT, we can reproduce the results of single-site DMFT. We set total system size $N_c+N_b\geq 10$, i.e., we choose 10 bath sites for 1-site case and 8 for all the rest cases. The self-consistency condition (\ref{selfconsistent}) is imposed on the imaginary frequency axis with a cutoff frequency larger than all the bath energies. We set an effective inverse temperature $\beta t=200$ which serves as a low-frequency cutoff. The number of bath parameters is reduced by considering the symmetry of the square and cubic lattices. All the energies are measured in units of $t$. 

\begin{figure}[tb]
\begin{center}
\includegraphics[width=6.0cm]{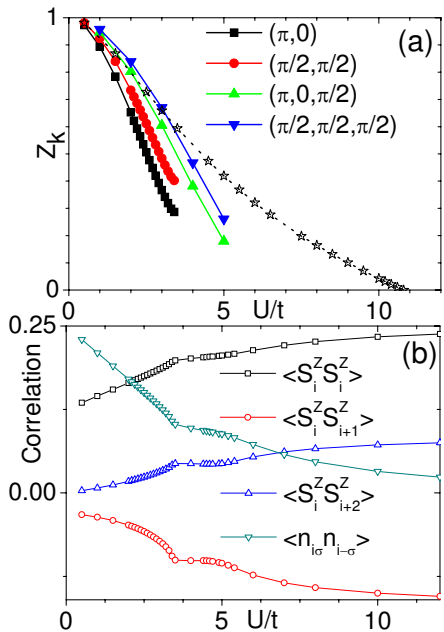}
\end{center}
\caption{(Color online) (a) Momentum and $U$ dependences of quasipaticle residue at Fermi surface for both cubic and square lattices obtained from 4-site CDMFT approach. The Lorentzian broadening factor is 0.1$t$. The 1-site CDMFT result for square lattice is  also shown by the dashed curve for comparison. (b) The local spin correlation (open square), nearest-neighbor spin correlation (open circle), next-nearest-neighbor spin correlation (up triangle) and double occupation number (down triangle) as a function of $U$ for square lattice obtained from 4-site CDMFT approach. Around $U_c$=3.5 (5), the central peak splits into twin peaks for square (cubic) lattice, leading to a sudden jump of $Z$ to vanish and a singular behavior in each correlation function.}
\label{Fig.2}
\end{figure}

We first show in Fig. \ref{Fig.1} the DOS for the half-filled square lattice. From Fig. \ref{Fig.1} (a) to (d), different choices of the cluster sizes up to 4 sites are presented. For single-site DMFT in Fig. \ref{Fig.1} (a), the central coherent peak is pinned at the Fermi level and becomes narrow as Hubbard $U$ increases. At $U=10t$, it is shown that the central peak is well seperated from the upper and lower Hubbard bands. As $U$ is larger than the transition point at $U \sim 11t$, the quasiparticle spectral weight at Fermi level vanishes and the system changes from a Fermi liquid state directly to an insulator. The quasiparticle residue 
\begin{equation}
Z_k=\left( 1-\frac{\partial \text{Re} \Sigma_{latt}\left( k,\omega \right) }{%
\partial \omega }| _{\omega \rightarrow 0} \right)^{-1}
\end{equation}
as function of increasing $U$ is shown in Fig. \ref{Fig.2} (a), where the lattice self-energy $\Sigma_{latt}$ is equal to the cluster self-energy for the 1-site case. Therefore $Z_k$ is $k$-independent. The quasiparticle residue continuously vanishes at the MIT which confirms the scenario of the MIT of DMFT\cite{Georges}. However, as the cluster size become larger than 2, we will show later that the situation dramatically changes due to the involvement of nonlocal correlations. 

Figure \ref{Fig.1} (b) to (d) show that, while broad central peaks still exist for smaller $U$, they are suppressed as new twin peaks appear just below and above the Fermi level when $U$ becomes larger (see $U=4t$ results calculated by 2-/4-site CDMFT and $U=6.4t$ by 3-site one). Here we introduce a Lorentzian broadening factor $\eta=0.1t$ to mimic the finite energy resolution in experiments or the finite temperature effect. The observed formation of a pseudogap prior to the Mott transition suggested in the present result offers a very different scenario of the MIT from that of DMFT\cite{Georges}. With further increasing value of $U$, the new twin peaks are separated further and eventually a Mott gap smoothly opens (see $U=8t$ ($7t$) for 2-/3-(4-) site CDMFT respectively). 

\begin{figure}[tb]
\begin{center}
\includegraphics[width=8.5cm]{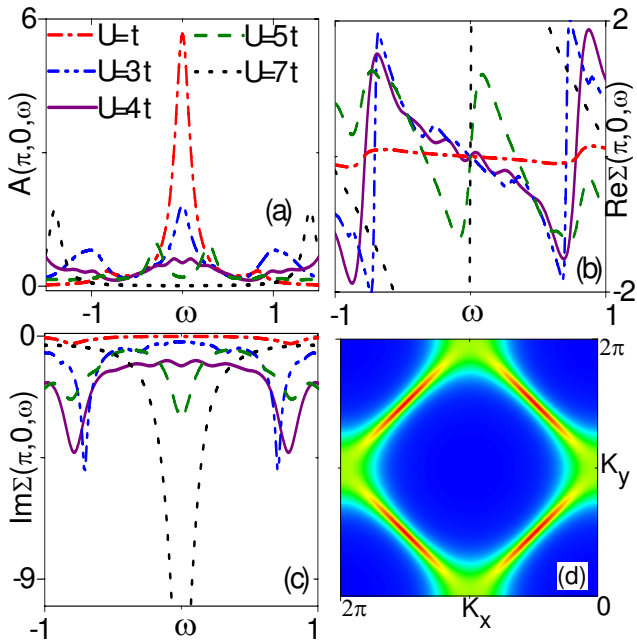}
\end{center}
\caption{(Color online) The spectral density $A \left( k, \omega \right)$ (a), real Re$\Sigma \left( k, \omega \right)$ (b), and imaginary Im$\Sigma \left( k, \omega\right)$ (c) parts of the self-energy at the Fermi surface $X$ point $k= \left( \pi,0 \right)$ for a 4-site cluster at various values of $U$ on square lattice. As U increases, a pseudogap develops in $A \left( k, \omega \right)$. A positive slope in Re$\Sigma \left( k, \omega \right)$ and a local minimum in Im$\Sigma \left( k, \omega\right)$ at $\omega=0$ indicate the breakdown of the Fermi-liquid behavior.(d) Density plot of $A(k,0)$ on square lattice in the metallic region at $U$=3.3$t$.The Lorentzian broadening factor is 0.1$t$ }
\label{Fig.3}
\end{figure}

The spectral density $A \left( k, \omega \right)$, the real Re$\Sigma \left( k, \omega \right)$ and imaginary Im$\Sigma \left( k, \omega\right)$ parts of the self-energy are shown in Fig. \ref{Fig.3} (a), (b), and (c), respectively, at the Fermi surface $X$ point $k= \left( \pi,0 \right)$ for a 4-site cluster at various values of $U$ on square lattice. The lattice self-energy $\Sigma_{latt}$ is derived from the cluster self-energy as\cite{Kotliar2,Parcollet,Biroli}
\begin{eqnarray}
&&\Sigma _{latt}\left( k\right) =\frac 14\sum_{i=1}^4\Sigma _{ii} \nonumber \\
&&+\frac 12[\Sigma _{14}\cos \left( k_x+k_y\right) +\Sigma _{23}\cos \left(
k_x-k_y\right)   \nonumber \\
&&+\left( \Sigma _{12}+\Sigma _{34}\right) \cos k_x+\left( \Sigma
_{24}+\Sigma _{13}\right) \cos k_y].
\end{eqnarray}
At the $X$ point, the slopes of Re$\Sigma \left( k, \omega \right)$ are positive at $U=4t$ and $5t$ , where we observed a formation of pseudogap in the $A \left( k, \omega \right)$. Simultaneously, Im$\Sigma \left( k, \omega\right)$ show local minima at $\omega =0$, which indicates a non-Fermi-liquid behavior. These results signal the appearance of two new solutions in the quasiparticle equation $\omega-\epsilon_{k}-\text{Re}\Sigma \left( k, \omega\right)$ in addition to the strongly damped solution at $\omega =0$ which is also present in the noninteracting system. This is the consequence of the involvement of the short-range nonlocal correlations. We find that, with increasing $U$, not only the local magnetic moment is enhanced and the double occupation number decreases but also the nonlocal magnetic correlations within the cluster are enhanced gradually as shown in Fig. \ref{Fig.2} (b). The nearest-neighbor spin-spin correlation is always negative so that the spin correlation is antiferromagnetic. It should be noted that the pseudogap appears even for the present cluster size where spin correlations are allowed to extend at most only up to the cluster size. 

The spectral density at the Fermi surface point $k=\left(\pi/2, \pi/2 \right)$ (not shown) indicates that the pseudogap appears more or less simultaneously for all the $k$ points at the Fermi surface for our limited cluster size up to 4 sites. The more or less simultaneous pseudogap opening is presumably due to the smallness of the cluster which cannot allow for the nonlocal correlations larger than the cluster size. It is plausible that the pseudogap opens at different $U$ for different $k$ points for a sufficiently large cluster. In the present results, we find that the separation of the two new poles at $k=\left(\pi/2, \pi/2 \right)$ is always smaller than that at $k=\left(\pi, 0 \right)$ as $U$ increases. Such a differentiation among different $k$ points is already seen even in the Fermi-liquid region. In Fig. \ref{Fig.3} (d), we plot the spectral intensity $A(k,\omega)$ at $\omega$=0 in the Brillouin zone at $U/t$=3.3. Cold (hot) region is visible around ($\pi/2$,$\pi/2$) (($\pi$,0)), implying a differentiation of MIT depending on the Fermi surface position. Near the MIT, the differentiation would make compatibly a gap in the hot region with a retained spectral weight in the cold region. For even larger $U=7t$, a large separation of the new twin poles at each $k$ point with a strong peak in Im$\Sigma \left( k, \omega\right)$ at the Fermi level leads to a large gap in $A \left( k, \omega \right)$. For smaller $U=t$ and $3t$, a Fermi liquid still survives.

From Fig. \ref{Fig.1}, we find that a gap opens at a smaller $U/t$ for 2-, 3- and 4-site CDMFT than for the 1-site CDMFT. This at least partially originates from a cooperative role of short-range spin correlation for the gap widening. The Mott transition occurs at substantially lower $U$ if the spatial antiferromagnetic  correlation is taken into account.  This indicates that the Mott mechanism of the gap opening at the metal-insulator transition is cooperatively enhanced by the Slater mechanism ascribed to the antiferromagnetic order.

As is seen from Fig. \ref{Fig.1}, there still exists a small difference between the results of 3 sites (odd number) and 2-/4 sites (even number). The central peak persists to a larger value of $U$ for the odd than for even sites. This may be due to the fact that the sharp coherent peak is associated with residual entropy arising from the unpaired spin (spin doublet structure) induced in odd number of sites as is expected in 1- and 3-site clusters.

\begin{figure}[tb]
\begin{center}
\includegraphics[width=6.0cm]{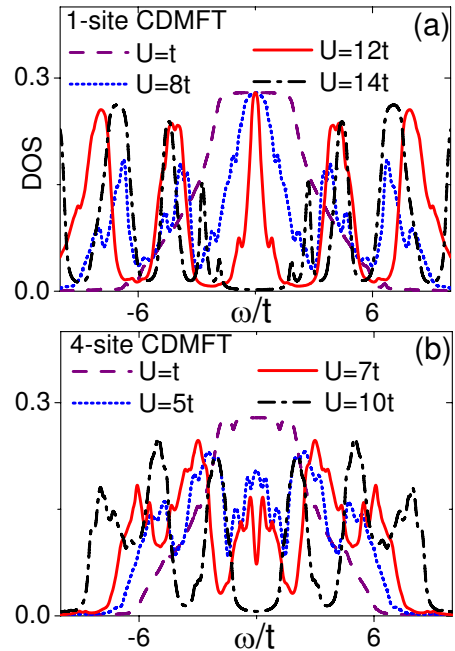}
\end{center}
\caption{(Color online) (a) and (b) are DOS of Hubbard model on half-filled cubic lattice, obtained from CDMFT approach with cluster size up to 4. (a) is the 1-site CDMFT which reproduce the results of single-site DMFT. (b) is 4-site CDMFT. The central peak splits into a twin-peak structure around Fermi level similarly to the square lattice. Here we take the broadening factor 0.1$t$.}
\label{Fig.4}
\end{figure}

In Fig. \ref{Fig.2} (a) we also show the quasiparticle residue on the Fermi surface with increasing $U$ obtained from the 4-site CDMFT approach. The most prominent difference from the single-site DMFT where the quasiparticle residue at the Fermi level vanishes continuously toward the transition point is that it decreases but remains at a nonzero value until new twin poles appear in the vicinity of the Fermi level at each $k$ point. Then the MIT is characterized by escape of the quasiparticle poles away from the Fermi level resulting in a discontinuous jump of $Z$ to zero, instead of the complete transfer of spectral weight from the well-seperated central peak to the upper and lower Hubbard bands. This is consistent with the recent study by the correlator projection theory\cite{Hanasaki}. From the CDMFT approach, the quasiparticle residue becomes $k$-dependent along the Fermi surface. The residue increases monotonically from $\left( \pi ,0\right) $ to $\left( \pi /2,\pi /2\right) $ at a given $U$. Therefore, we only show two $k$ points at Fermi surface in Fig. \ref{Fig.2} (a). In Fig. \ref{Fig.2} (b), no jump is detected in all the correlations, indicating that the splitting of the quasiparticle poles at the Fermi surface into the twin poles in the vicinity of it is continuous. 

It is interesting to see whether the results qualitatively change if the dimensionality increases to three. We show in Fig. \ref{Fig.4} the DOS for half-filled cubic lattice by using 1-site and 4-site CDMFT approaches. No qualitative difference is found between square and cubic lattices. A twin-peak structure around the Fermi level is found by 4-site CDMFT approach as $U$ increases (see $U=7t$). The $k$-dependent quasiparticle residue, as shown in Fig. \ref{Fig.2} (a), remains nonzero at the proximity of the point where the Fermi-liquid behavior breaks down while the value is smaller than that of square lattice indicating a crossover to infinite dimension where the quasiparticle residue at Fermi level should continuously vanish. 

The results on the cubic lattice are relevant to the recent high-resolution bulk-sensitive PES of SrVO$_3$ and CaVO$_3$. At finite temperature $T=6K$ and with the energy resolution 8meV, a peak around 0.2eV is observed with a pseudogap structure in qualitative but essential agreement with our results near the Mott transition. 

In summary, we have shown that the MIT scenario obtained by CDMFT with cluster size larger than 2 is quite different from that of the single-site DMFT. With the increase of the interaction $U$ in the CDMFT, the quasiparticle peak becomes continuously split into twin peaks moving away from the Fermi level and leaving a pseudogap at the Fermi level on both the square and cubic lattices. With further increase of $U$, a full gap opens gradually from hot to cold spots. As a consequence, the quasiparticle weight remains $k$-dependent but nonzero until the splitting of the central peak which leads to a sudden jump of $Z$ to vanish. This is in contrast to the single-site DMFT, where it is $k$-independent and vanishes continuously at the MIT point. We find that the transition is not driven by a uniform and continuous vanishing of quasiparticle weight.  The transition is rather driven by the escape of quasiparticle poles away from the Fermi surface.  The momentum differentiation of the escaping process leaves a truncated structure of the Fermi surface such as the "arc"-type structure.  This leads to a topological change of the Fermi surface at the transition to the insulator. The present unravelled mechanism of MIT characterized by the topological and momentum-resolved change in the Fermi surface indicates a route to the Mott transition without an isotropic divergence of the electronic effective mass and opens a way to study profound implications expected for the proximity from the MIT.

We thank S. Shin for illuminating discussions on the experimental results. This work is supported by Grant-in-Aids for Scientific Research on Priority Areas under the grant numbers 17071003 and 16076212 from MEXT, Japan. The authors also thank supports from the Supercomputer Center, ISSP, University of Tokyo.

\end{document}